\documentclass[a4paper,twocolumn,english,aps,prl,superscriptaddress]{revtex4}
\newcommand{\Tr}{\mbox{Tr}}
\usepackage[T1]{fontenc}
\usepackage[latin1]{inputenc}
\usepackage{float}
\usepackage{graphicx,color}
\usepackage{bm}
\usepackage{bbm}
\usepackage{amsmath}
\usepackage{color}
\usepackage{amssymb}
\makeatletter

\usepackage{babel}
\makeatother
\begin{document}

\title{Quantum statistics of four-wave mixing by a non-linear resonant microcavity}

\author{Y. Sherkunov}
\address{Physics Department, Lancaster University, Lancaster, LA1 4YB,
UK}
\author{David M. Whittaker}
\address{Department of Physics and Astronomy, University of Sheffield, Sheffield S3 7RH, UK}

\author{Henning Schomerus}
\address{Physics Department, Lancaster University, Lancaster, LA1 4YB,
UK}
\author{Vladimir Fal'ko}
\address{Physics Department, Lancaster University, Lancaster, LA1 4YB,
UK}

\begin{abstract}
We analyse the correlation and spectral properties of two-photon states resonantly transmitted by a non-linear optical microcavity. We trace the correlation properties of transmitted two-photon states to the decay spectrum of multi-photon resonances in the non-linear microcavity.       
\end{abstract}
\maketitle

Generation and controlled propagation of two-photon states in optical circuits represent a new challenge in quantum optics -- the next step from the earlier-developed single-photon sources. Photon pairs have been successfully produced using semiconductor quantum dots \cite{Stevenson06,Akopian06,Salter10,Bennett10,Stevenson12} and by coupling superconducting qubits to high-quality microwave resonators \cite{Wallraff, microwave, Hofheinz08, Hofheinz09, Mariantoni11, Devoret13, Yin13}. However, the controlled propagation of correlated photon pairs across optical circuits remains an unexplored territory.

Here, we propose a theory describing the transmission of two-photon states in optical circuits where junctions are non-linear optical or microwave resonators  \cite{Nozaki,Ngueng,Verger,Carusotto}. In such systems, the interaction between photons inside the cavity both shifts the conditions for their resonant transmission and leads to four-wave-mixing -- a redistribution of photon energies in the transmitted pair. We show that the four-wave-mixing by the non-linear cavity results in  spectral correlations of the transmitted two-photon states, prescribed by the decay of multi-photon resonances in a non-linear cavity, and we analyse quantum statistics of few-photon states emitted by a non-linear resonator under the action of a weak classical pump.

\begin{figure}[h]
\includegraphics[width=0.45\textwidth]{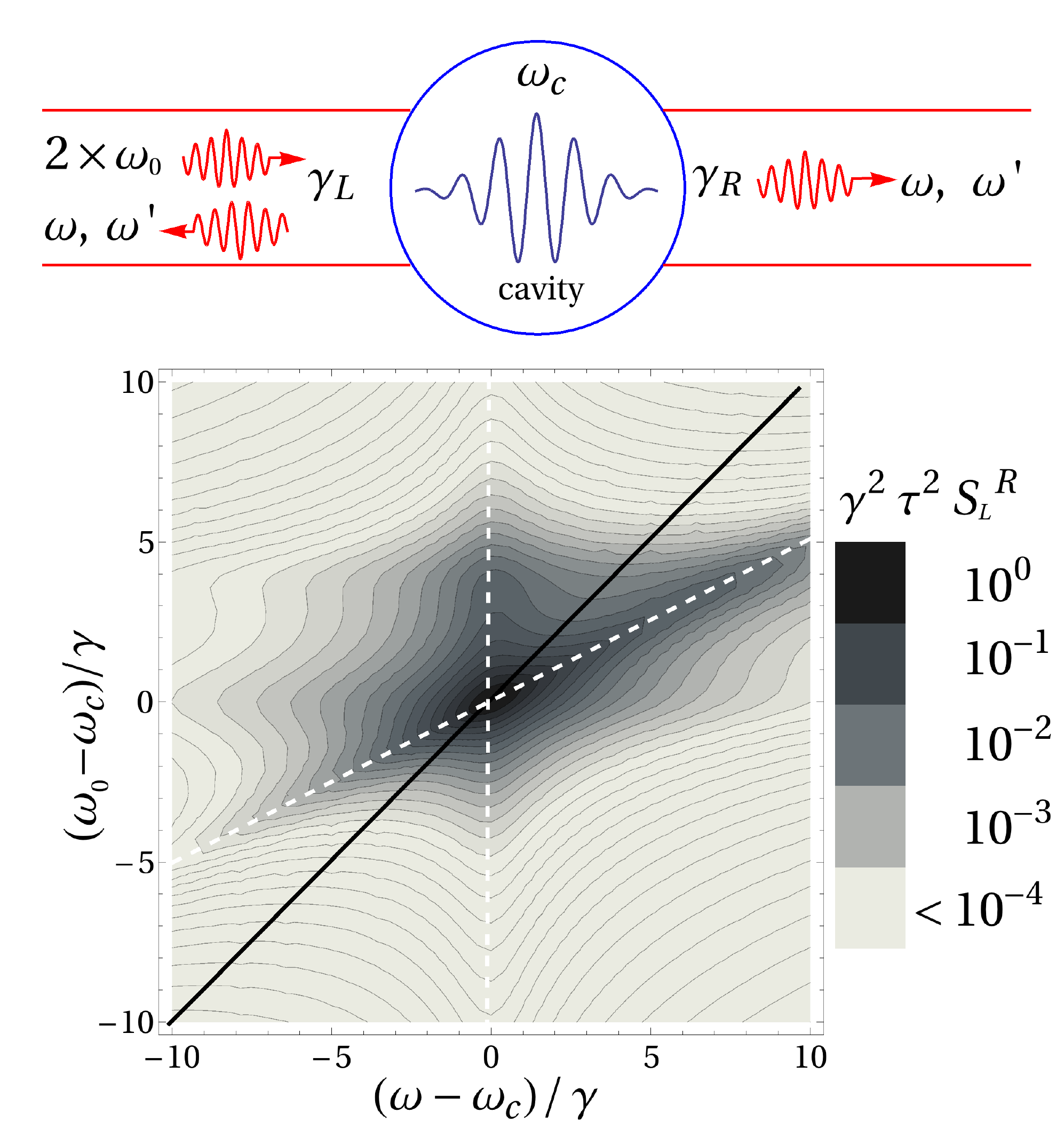}
\caption{Two-photon transmission across a non-linear resonator junction in an optical circuit. Top: Schematic of an optical cavity resonator connecting two waveguides.  Bottom: Contour plot of spectral function $S_{L}^{R}$ characterising spectral properties of transmitted two-photon states, as a function of frequencies $\omega$ of the outgoing photons and $\omega_0$ of the incident monochromatic photons, for  $\gamma_L=\gamma_R=\gamma$ and non-linear coupling $u/\gamma=4$. The single-photon transmission peak is indicated by the black line at $\omega=\omega_0$; maxima of $S_{L}^{R}$ at $\omega=\omega_c$ and $\omega=2\omega_0-\omega_c$ are traced using the white dashed lines.   }
\label{Fig1}
\end{figure}

The key element of the optical circuits discussed in this paper is sketched in Fig. \ref{Fig1}. It consists of a non-linear microcavity, with  resonance frequency $\omega_c$ and decay broadening $\gamma=(\gamma_L+\gamma_R)/2=\omega_c/Q$ ($Q$ stands for the quality factor) determined by couplings $\gamma_{L(R)}$ to the left(right) waveguides. The interaction of two incoming photons of frequency $\omega_0$ inside the cavity generates two outgoing photons with frequencies $\omega$ and $\omega '= 2\omega_0-\omega$ ($\omega$, $\omega'\approx \omega_0$). As a function of the frequency of an outgoing photon $\omega$, the characteristic spectral density of the transmitted photons is illustrated by the colour-scale plot in Fig. \ref{Fig1}. The spectral density shows distinctive maxima at $\omega_c$ and $2\omega_0-\omega_c$. These statements also apply to a nearly monochromatic pair in a two-photon pulse with a temporal extent of $\tau \gg \gamma^{-1}$. Figure \ref{Fig1} also shows that the four-wave mixing is most efficient either when the incoming photons' frequency resonates with the empty cavity mode, or when the total energy $2\hbar\omega_0$ of the incoming pair coincides with the energy of the interacting photon pair inside the cavity $2\hbar(\omega_c+u)$, where $2\hbar u$ is the interaction energy. Here, we show that spectral properties of the transmitted photon pair  are similar over a broad range of cavity and incoming pulse parameters. 
The spectral density in Fig. \ref{Fig1} illustrates  spectral properties of the resonant four-wave-mixing, where, as we show below, the two-photon spectral function can be factorised into an $\omega_0$-dependent probability to create a two-photon state in the cavity and the spectral density of the following decay of such a state due to the photon escape into the waveguides.

The evolution of photons transmitted by the non-linear resonator sketched in Fig. \ref{Fig1}  is modelled using the Hamiltonian,

\begin{eqnarray}
H&=&H_0+H_u+H_L+H_R,\label{Ham}\\
H_0&=&\omega_c \alpha^{\dagger}\alpha,\nonumber\\
H_u&=&u\alpha^{\dagger}\alpha^{\dagger}\alpha\alpha,\nonumber\\ 
H_{L}&=&\int dx[\beta_L^{\dagger}(x)\left(-i\frac{\partial}{\partial x}\right)\beta_L(x)\nonumber\\
&+&\sqrt{\gamma_L}\delta(x)(\beta_L^{\dagger}(x)\alpha+\alpha^{\dagger}\beta_L(x))],\nonumber\\
H_{R}&=&\int dx[\beta_R^{\dagger}(x)\left(-i\frac{\partial}{\partial x}\right)\beta_R(x)\nonumber\\
&+&\sqrt{\gamma_R}\delta(x)(\beta_R^{\dagger}(x)\alpha+\alpha^{\dagger}\beta_R(x))],\nonumber
\end{eqnarray}
where $\alpha$/$\alpha^{\dagger}$ is the annihilation/creation operator of the resonant cavity mode with energy $\omega_c$ (here, fundamental constants are set to $c=1$ and $\hbar =1$). The non-linear term, $H_u$, in Eq. (\ref{Ham}) with coupling constant $u$, leads to photon-photon interaction, shifting the two-photon state energy by $2u$. This non-linearity is generic for inversion-symmetric media; in particular, it can be the result of the repulsion between exciton-polaritons in a microcavity with a large Rabi frequency \cite{Nozaki,Ngueng,Verger,Carusotto}. The resonant cavity mode is coupled to the photons in the left (L) and right (R) semi-infinite waveguides, described by annihilation operators in the real-space representation, $\beta_{L(R)}(x)=\frac{1}{\sqrt{s}}\sum_{k>0} e^{ikx}\beta_{L(R)}(k)$ ($k$ is the wave-number and $s$ is the length of the waveguide), and $\beta_{L(R)}(x)$ are defined to describe incoming photons for $x<0$ and outgoing photon for $x>0$ (irrespective of the side of the resonator). The transformation of propagating waves into the cavity mode (and their reflection from the cavity) takes place at $x=0$. 

\begin{figure}[h]
\includegraphics[width=0.4\textwidth]{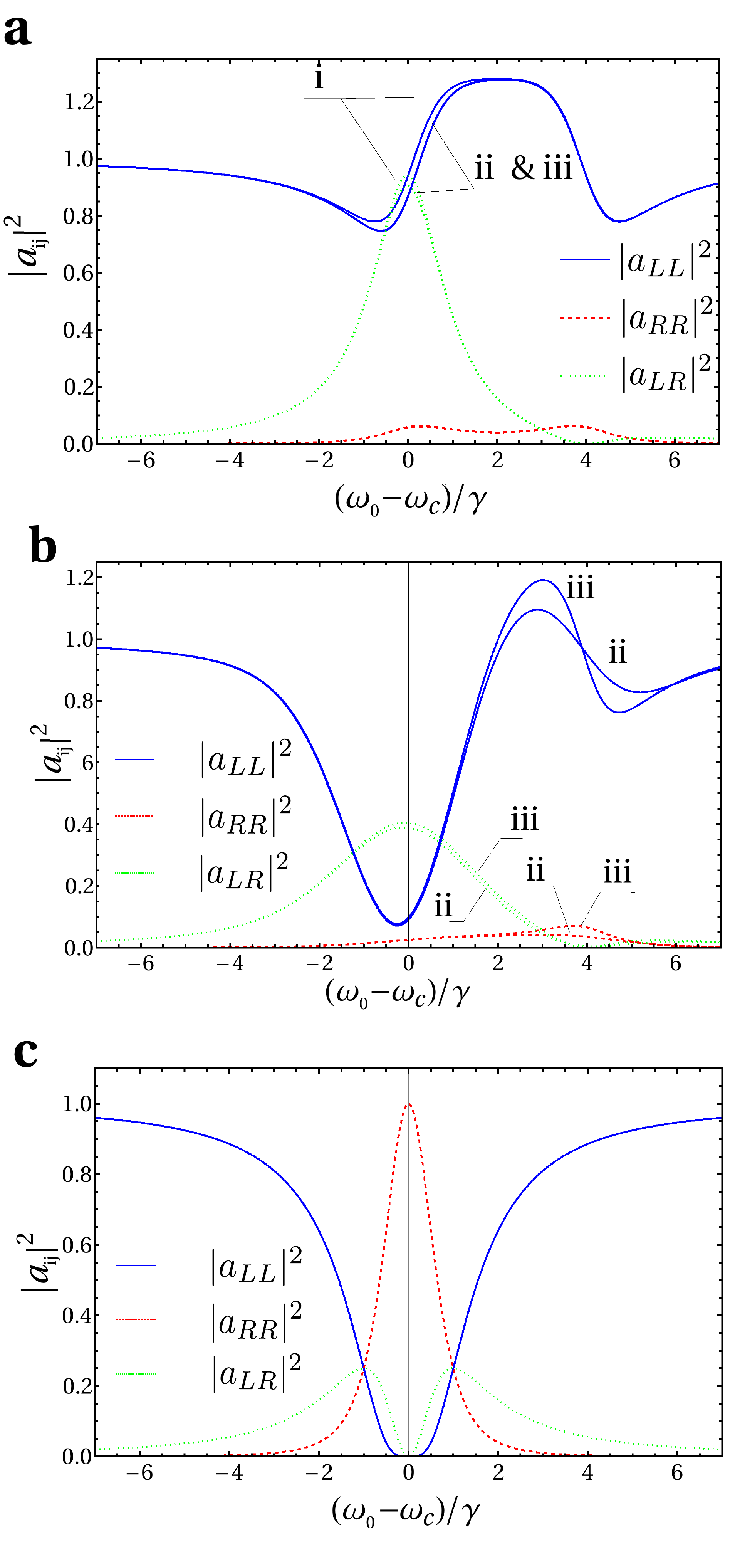}\\
\caption{Transmission, $|a_{LL}|^2$, reflection, $|a_{RR}|^2$, and transmission-reflection, $|a_{LR}|^2$  amplitudes of two-photon states in various parametric regimes. 
(a) Non-linear cavity with $\tau\gamma=7$ ($\gamma_L=\gamma_R=\gamma$) and $u=4\gamma$. Frequency dependence of two-photon amplitudes $|a_{ij}|^2$ at $x_1=x_2$ for (i) a monochromatic incident wave, (ii) uncorrelated initial pair of photons, and (iii) a correlated Gaussian initial state (see Table \ref{table:table1}). (b) Same as (a), but for $\tau\gamma=1$. (c) Transmission by a linear cavity ($u=0$). }
\label{Fig2}
\end{figure}

The solution for the two-photon transmission/reflection problem is obtained by studying the evolution of the two-photon states $|\psi(t)\rangle$, governed by the Schr{\"o}dinger equation, $i\partial_t|\psi(t)\rangle =H|\psi(t)\rangle$. We match the initial state to the incident photon pair in the left waveguide,  
\begin{eqnarray}
|\psi\rangle_{t\rightarrow -\infty}=\frac{e^{i\omega_0(x_1+x_2-2t)}}{\tau\sqrt{2}} f(x_1,x_2,t)\beta^{\dagger}_L(x_1)\beta^{\dagger}_L(x_2)|0\rangle,
\label{initialcond}
\end{eqnarray}
where the two-photon pulse envelope $f(x_1,x_2,t)$ has a temporal extent $\tau$ and is modelled using the scenarios  listed in Table \ref{table:table1}. The projection of the transmitted/reflected photon pairs onto the outgoing states in the left/right leads ($i=$L/R) is described using 
\begin{eqnarray}
|\psi\rangle_{t\rightarrow \infty}=\frac{e^{i\omega_0(x_1+x_2-2t)}}{\tau\sqrt{2}}a_{ij}(x_1,x_2,t)\beta^{\dagger}_i(x_1)\beta^{\dagger}_j(x_2)|0\rangle,
\label{psiscat}
\end{eqnarray}
where, for a monochromatic photon pair ($\tau\rightarrow\infty$), the amplitudes $a_{LL}$ ($a_{RR}$) of obtaining two photons in the left(right) waveguide and $a_{LR}$ of finding one photon in each waveguide take the form (see Appendix A): 
\begin{eqnarray}
a_{ij} &=& c_i c_j - \frac{u\gamma_L\sqrt{\gamma_i\gamma_j} e^{[i(\omega_0-\omega_c)-\gamma]|x_1-x_2|}}{(\omega_0-\omega_c+i\gamma)^2(\omega_0-\omega_c-u+i\gamma)}, \label{r2}\\
c_L&=&-\frac{\omega_0-\omega_c+i(\gamma_R-\gamma_L)/2}{\omega_0-\omega_c+i\gamma}, \nonumber \\
c_R&=&\frac{i\sqrt{\gamma_L\gamma_R}}{\omega_0-\omega_c+i\gamma}. \nonumber
\end{eqnarray}
Here, $c_i$ is a single-photon reflection (L) and transmission (R) amplitude in the circuit with a resonant cavity. The first term in Eq. (\ref{r2}) accounts for independent scattering of photons. The second term describes the correlated two-photon states formed due to the photon-photon interaction inside the cavity. Spectral properties of the amplitudes $a_{ij}$, calculated for various finite pulse widths, are illustrated in Fig. \ref{Fig2}, where we show that the two-photon transmission by the non-linear cavity is strongly affected by the photon-photon interaction, but does not depend much on the detailed spectral shape of the two-photon pulse. The examples in Fig. \ref{Fig2}(a,b) demonstrate that the interaction of two photons  inside the cavity produces their anti-bunching for $\omega_0=\omega_c$, but also generates an additional transmission peak at  $\omega_0=\omega_c+u$ corresponding to the formation of resonant two-photon states inside the cavity. Note that in the limit $u\rightarrow \infty$, the model in Eq. (\ref{Ham}) corresponds to the photon blockade regime studied in Ref. \cite{Shen} for an optical junction consisting of a two-level system, and, in this limit, Eq. (\ref{r2}) reproduces the photon anti-bunching predicted in Ref. \cite{Shen}. 

\begin{table}[h]
\caption {Incident states}
\begin{tabular}{cc}
\hline\hline
State & $f(x_1,x_2,t)$\\
\hline 
(i) Monochromatic & 1\\
(ii) Uncorrelated Gaussian & $\pi^{-1/2}e^{-[(x_1-t)^2+((x_2-t)^2]/2\tau^2}$\\
(iii) Correlated Gaussian & $\pi^{-1/4}e^{-(x_1-x_2)^2/2\tau^2}$\\
\hline\hline
\end{tabular}
\label{table:table1}
\end{table}

To characterise the spectral properties of the photon pair transmitted/reflected by the non-linear cavity into a waveguide $i$ from a pulse arriving over a temporal interval $\tau$ (see Table \ref{table:table1}) in  waveguide $j$, we consider the spectral density of photons detected at a distant position $x_0\rightarrow \infty$ in waveguide $i$, \footnote{We assume a linear dispersion of the wave guide, which implies the following relation for the spatial and temporal variables $x=ct$, where $c$ is the speed of light .}  
\begin{eqnarray}
S_{j}^{i}(\omega) \equiv \int_{-\infty}^{\infty}\frac{dt}{2\pi}e^{-i\omega t} \langle\beta_{i}^{\dagger}(x_0)\beta_{i}(x_0-t)\rangle_{j}. \nonumber 
\end{eqnarray}
We find that the latter can be written in the form
\begin{eqnarray}
S_{j}^{i}(\omega)= \frac{\gamma_i^2\gamma_{j}}{\gamma^3}F(\omega_0)\tilde{F}(\omega)+I_{j}^{i}\delta (\omega-\omega_0),
\label{S}
\end{eqnarray}
where the first term describes the photon-photon correlations introduced by the interaction between two photons simultaneously appearing inside the cavity, whereas the second term takes into account elastic (uncorrelated) single-photon transfer.
In Eq. (\ref{S}), the factor
\begin{eqnarray}
F(\omega_0)&=&\frac{2\gamma^2/\pi \tau^2}{[(\omega_0-\omega_c)^2+\gamma^2][(\omega_0-\omega_c-u)^2+\gamma^2]} \nonumber
\end{eqnarray}
is the probability to form the correlated two-photon state inside the cavity, whereas the factor  
\begin{eqnarray}
\tilde{F}(\omega)&=&\frac{4u^2\gamma^2}{[(\omega-\omega_c)^2+\gamma^2][(\omega-2\omega_0+\omega_c)^2+\gamma^2]}\nonumber
\end{eqnarray} 
describes the spectral density of the photons emitted to the left/right waveguide. 
Finally, single-photon transfer is described by $I_{L}^{L}=I_{R}^{R}=2|c_L|^2/\tau$ and  $I_{L}^{R}=I_{R}^{L}=2|c_R|^2/\tau$.

\begin{figure}[h]
\includegraphics[width=0.4\textwidth]{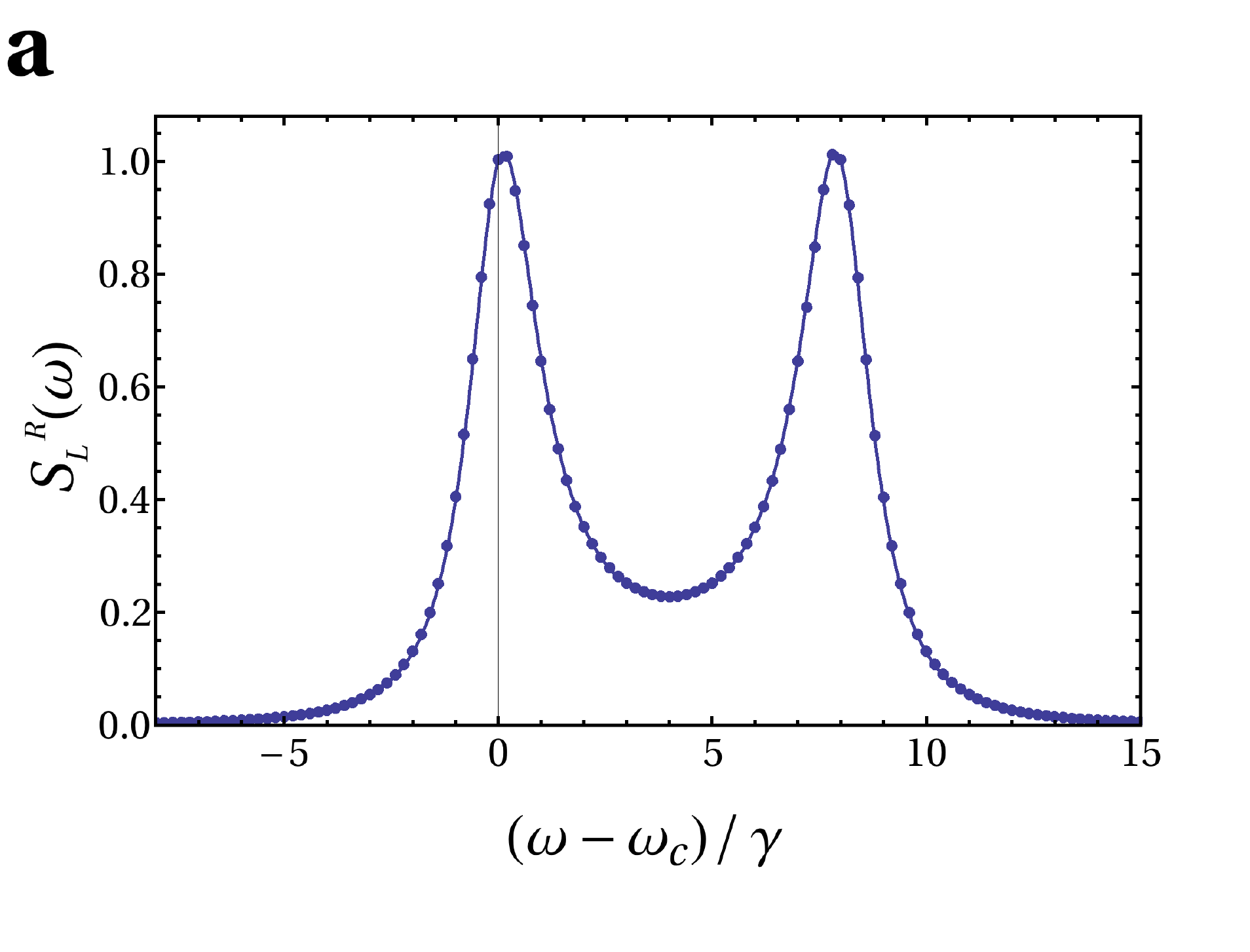}\\
\includegraphics[width=0.4\textwidth]{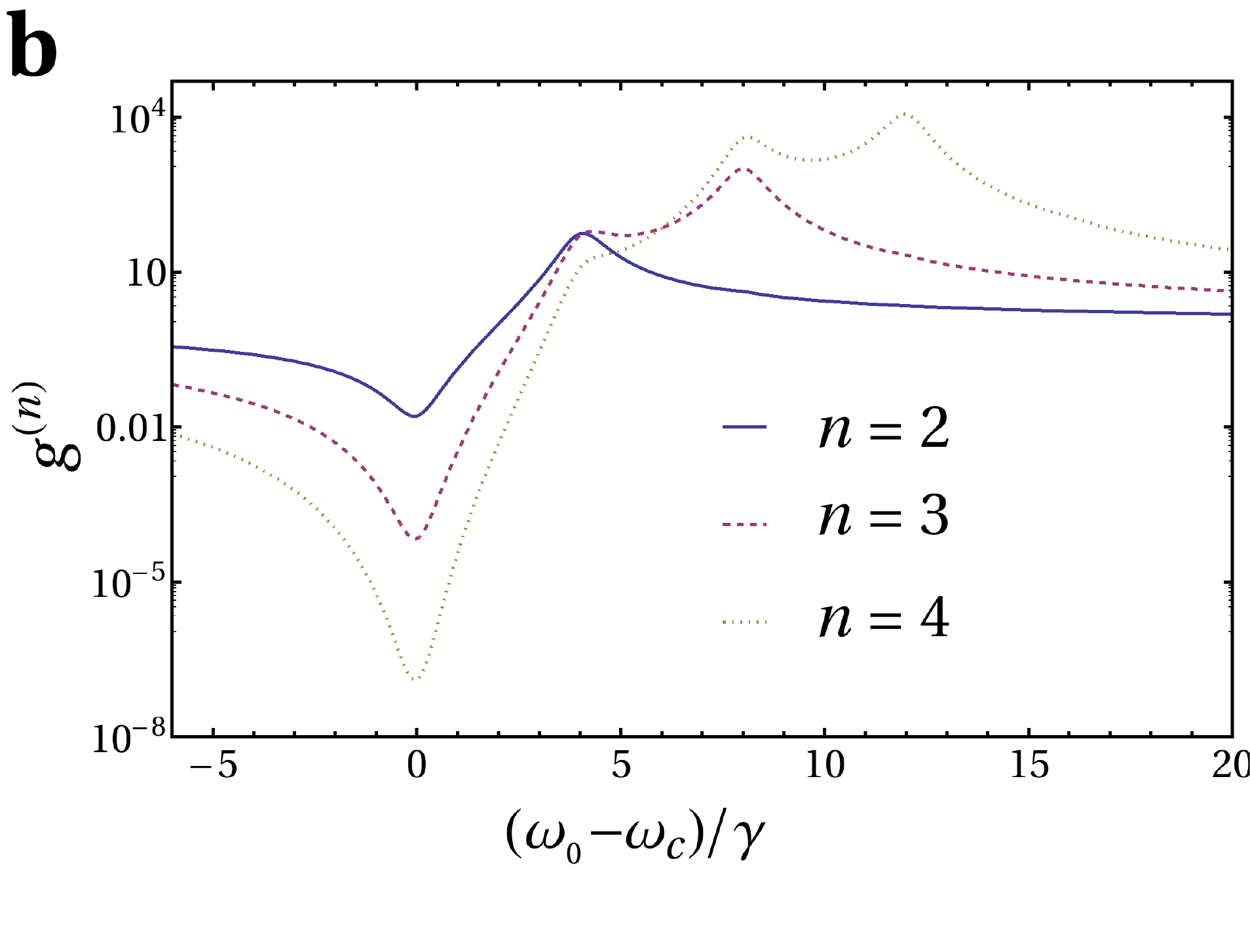}
\caption{Generation of few-photon states in a non-linear resonator excited by a classical field.
(a) Spectral function $S_L^R(\omega)$ (normalized to its maximum value) for a cavity excited by a classical field source for $\gamma_L=\gamma_R=\gamma$, $(\omega_0-\omega_c)/\gamma=-4$, $u/\gamma=4$, $\sqrt{b/\gamma\tau}=0.1$, ($|\langle \alpha\rangle|^2=5.9\times 10^{-4}$). Dots represent the results of numerical calculations, and solid line a fit to the spectral density described by Eq. (\ref{S}). 
(b) Few-photon correlation function $g^{(n)}(\omega_0)$ for two-, three-, and four-photon states emitted to the right waveguide by a non-linear optical resonator excited by a classical field arriving from the left waveguide. }
\label{pumpspectr}
\end{figure}

The weak dependence of the correlation properties of scattered photons on the correlation properties of the initial state seen in Fig. \ref{Fig2} indicates that the result in Eq. (\ref{S}) is applicable to describe correlated photon pairs produced by incident two-photon states with different correlation properties, provided that $\tau\gamma \gg 1$. Moreover, the two-photon spectral correlations, described by Eq. (\ref{S}), emerge even in the case of initial state with different quantum statistics.  To demonstrate this, we studied the emission spectrum by a non-linear resonator excited by a coherent state $|\rm{in}\rangle=e^{-|b|^2/2}e^{b\beta^{\dagger}_{\omega_0}}|0\rangle$ incident from one of the waveguides, e.g, $j=L$. Here, $|b|$ characterizes the amplitude $\langle \rm{in}|\beta|\rm{in}\rangle$ of the incoming field pulse.  In this analysis, we exploit the possibility to relate the spectral function $S_L^R$ to the correlation function of the cavity resonance mode, as established in Ref. \cite{WallsMilburn}, 
\begin{eqnarray}
S_L^R(\omega)=\gamma_R\int_{-\infty}^{\infty} \frac{dt}{2\pi}e^{-i\omega t}\Tr \alpha^{\dagger}(t)U(t,0)\alpha(0)\rho(0), \label{SSSS}
\end{eqnarray}
where the operators $\alpha$, $\alpha^{\dagger}$ are used in the interaction representation, and $\rho$ is the reduced density matrix obtained from the full density matrix describing the system by tracing out the degrees of freedom associated with the quantum waveguide modes. The time dependence of the reduced density matrix $\rho$ obeys the Lindblad equation:
\begin{eqnarray}
\partial_t\rho&=&-i[H_s,\rho]+\gamma(2\alpha\rho \alpha^{\dagger}-\alpha^{\dagger}\alpha\rho-\rho \alpha^{\dagger}\alpha),\nonumber\\
\rho(t')&=&U(t',t)\rho(t),\nonumber\\
H_s&=&\omega_c \alpha^{\dagger}\alpha+u\alpha^{\dagger}\alpha^{\dagger}\alpha\alpha\nonumber\\
&+&\left[\sqrt{\gamma_L/\tau}be^{-i\omega_0t}\alpha^{\dagger}+h.c. \right], \label{Lind}
\end{eqnarray}    
where $U(t',t)$ is the evolution super operator of the reduced density matrix. When $U(t,t')$ is found using perturbation theory  (see  Appendix B) for a weak non-linearity, $u \ll \gamma$, this results in the spectral density described by Eq. (\ref{S}), but rescaled as 
$S_L^R \rightarrow |b|^4S_L^R$. 
A typical result of the numerical solution of Eq. (\ref{Lind}) for a strong non-linearity but weak pumping ($\langle \alpha^{\dagger}\alpha\rangle\ll 1$) is shown in Fig. \ref{pumpspectr}. It indicates that in the case of weak coherent pumping the spectrum of emitted photons  is determined by two-photon correlations since its density coincides with that described by Eq. (\ref{S}).  

In addition, we analyse the conditions for the generation of correlated few-photon states with two, three and four photons for a non-linear cavity excited by a classical field. The appearence of such states is quantified using the correlation function 
$$g^{(n)}=\langle(\beta_i^{\dagger})^n\beta_i^n\rangle/\langle \beta_i^{\dagger}\beta_i\rangle^n,$$
where  $\beta$-operators are taken at coinciding time.
By generalising the analysis in Ref. \cite{WallsMilburn}, this can be expressed in terms of the photon operators inside the cavity, 
$$g^{(n)}=\Tr[(\alpha^{\dagger})^n(\alpha)^n\rho]/\Tr [\alpha^{\dagger}\alpha]^n,$$
which can be determined from Eq. (\ref{Lind}). The results of numerical simulations for $n=2,3,4$ are shown in Fig. \ref{pumpspectr} b. Here, the second-order correlation function shows anti-bunching ($g^{(2)}<1$) for red-detuned pumping frequencies and bunching ($g^{(2)}>1$) for a blue-shifted frequencies, in agreement with \cite{Verger}. Also, the functions $g^{(n)}$ peak at pumping frequencies close to  $\omega_0=\omega_c+(n-1)u$, which correspond to the resonance conditions for a simultaneous excitation of $n$ photons inside the non-linear cavity. 

To summarise, we show that two-photon states resonantly transmitted by a non-linear optical microcavity display spectral correlations, which are almost independent of the correlation properties of the incoming photon pair and can be traced to the decay spectrum of multi-photon resonances in the cavity.  

\begin{acknowledgements} We thank B. Altshuler, J. Bloch, A. Cleland, Y. Pashkin, and M. Skolnick for useful discussions. This work was supported by EPSRC Programme Grant EP/J007544.
\end{acknowledgements}
\appendix
\section{Appendix A: Propagation of  two-photon states: Scattering approach}
 \renewcommand{\theequation}{A\arabic{equation}}
  % redefine the command that creates the equation no.
  \setcounter{equation}{0}  % reset counter 
 
A generic two-photon state $|\psi(t)\rangle$, whose evolution obeys
$i\frac{\partial}{\partial t}|\psi(t)\rangle=H|\psi(t)\rangle$,
can be written as  
\begin{eqnarray}
|\psi(t)\rangle&=&1/\sqrt{2}\left[\int dx_1 dx_2\psi_{LL}(x_1,x_2,t)\beta_L^{\dagger}(x_1)\beta_L^{\dagger}(x_2)|0\rangle \right. \nonumber\\
&+& \left.\int dx_1 dx_2\psi_{RR}(x_1,x_2,t)\beta_R^{\dagger}(x_1)\beta_R^{\dagger}(x_2)|0\rangle \right]\nonumber\\
&+&\sqrt{2}\int dx_1 dx_2\psi_{LR}(x_1,x_2,t)\beta_L^{\dagger}(x_1)\beta_R^{\dagger}(x_2)|0\rangle\nonumber\\
&+&\int dx\left[\psi_{LC}(x,t)\beta_L^{\dagger}(x)\alpha^{\dagger}+\psi_{RC}(x,t)\beta_R^{\dagger}(x)\alpha^{\dagger}\right]|0\rangle\nonumber\\
&+&1/\sqrt{2}\psi_{CC}(t)\alpha^{\dagger}\alpha^{\dagger}|0\rangle.
\end{eqnarray}
Here, $|0\rangle$ is the vacuum state, $\psi_{ij}$ are the wave functions describing photons in the left (right) waveguide ($i,j=L(R)$), or in the cavity ($i,j=C$).  This leads to the following system of simultaneous equations:
\begin{eqnarray}
&i&\left(\partial_{x_1}+\partial_{x_2}+\partial_{t}\right)\psi_{ij}(x_1,x_2,t)=1/\sqrt{2}\left[\kappa_i\delta(x_1)\psi_{jC}(x_2,t)\right.\nonumber\\
&+&\left.\kappa_j\delta(x_2)\psi_{iC}(x_1,t)\right],\nonumber\\
&i&\left(\partial_{x}+\partial_{t}+i\omega_{c}\right)\psi_{iC}(x,t)=\sqrt{2}\left[\kappa_i(\delta(x)\psi_{CC}(t)\right.\nonumber\\
&+&\left.\psi_{ii}(0,x,t))+\kappa_j\psi_{ij}(x,0,t)\right],i\neq j,\nonumber\\
&i&\left(\partial_{t}+i2(\omega_{c}+u)\right)\psi_{CC}(t)=\sqrt{2}\left[\kappa_L\psi_{LC}(0,t)\right.\nonumber\\
&+&\left.\kappa_R\psi_{RC}(0,t)\right].\nonumber
\end{eqnarray}

For the incident  state $|\psi\rangle_{t\rightarrow -\infty}=\frac{1}{\tau\sqrt{2}} e^{i\omega_0(x_1+x_2-2t)}f(x_1,x_2,t)\beta^{\dagger}_L(x_1)\beta^{\dagger}_L(x_2)|0\rangle$ (see Eq. (2)) and $\psi_{CC}(t\rightarrow -\infty)=\psi_{LC}(t\rightarrow -\infty)=\psi_{RC}(t\rightarrow -\infty)=0$,  we find that 
\begin{eqnarray}
\psi_{LL}(x_1,x_2,t)&=&\frac{e^{i\omega_0(x_1+x_2-2t)}}{\tau}\left[f(x_1,x_2,t)\theta(-x_1)\theta(-x_2)\right.\nonumber\\
&+&a_{LL}(x_1,x_2,t)\theta(x_1)\theta(x_2)\nonumber\\
&+&\left.c_L(\theta(x_1)\theta(-x_2)+\theta(-x_1)\theta(x_2))\right],\nonumber\\
\psi_{RR}(x_1,x_2,t)&=&\frac{e^{i\omega_0(x_1+x_2-2t)}}{\tau}a_{RR}(x_1,x_2,t)\theta(x_1)\theta(x_2),\nonumber\\
\psi_{LR}(x_1,x_2,t)&=&\frac{e^{i\omega_0(x_1+x_2-2t)}}{\tau}\left[a_{LR}(x_1,x_2,t)\theta(x_1)\theta(x_2)\right.\nonumber\\
&+&\left.c_R\theta(-x_1)\theta(x_2)\right],\label{psiscat}
\end{eqnarray}  
where $a_{LL}$, $a_{RR}$ and $a_{LR}$ are the amplitudes describing two reflected photons, two transmitted photons and one reflected and one transmitted photons respectively. In the case of monochromatic incident wave ($\tau\gamma \gg 1$) (i), they are given by:
\begin{eqnarray}
a_{ij}(x_1,x_2)&=&c_ic_j\nonumber\\
&-&\frac{u\gamma_L\sqrt{\gamma_i\gamma_j}\exp\{[i(\omega_0-\omega_c)-\gamma]|x_1-x_2|\}}{(\omega_0-\omega_c+i\gamma)^2(\omega_0-\omega_c-u+i\gamma)},\nonumber\\
c_L&=&-\frac{\omega_0-\omega_c+i(\gamma_R-\gamma_L)/2}{\omega_0-\omega_c+i\gamma},\nonumber\\
c_R&=&\frac{i\sqrt{\gamma_L\gamma_R}}{\omega_0-\omega_c+i\gamma}\nonumber.
\end{eqnarray}

\section{Appendix B: Classical pumping, weak interaction limit}
\renewcommand{\theequation}{B\arabic{equation}}
  % redefine the command that creates the equation no.
  \setcounter{equation}{0}  % reset counter 
In the case of weak non-linearity $\langle \alpha^{\dagger} \alpha\rangle u \ll\gamma$, Eq. (\ref{Lind}) can be solved analytically by expanding the density matrix over a coherent basis $|a\rangle$ so that $\rho=\int P(a)|a\rangle\langle a|d(\Re \{a\})d (\Im \{a\})$ ($P$-representation) \cite{WallsMilburn} and  converting  into a Fokker-Planck equation with the help of the operator equivalence relations: 
$\alpha\rho \rightarrow a P(a)$,
$\alpha^{\dagger}\rho\rightarrow \left(a^*-\partial_{a}\right)P(a)$,
$\rho \alpha\rightarrow \left(a-\partial_{a^*}\right)P(a)$,
$\rho \alpha^{\dagger}\rightarrow a^*P(a)$. Switching to the rotating frame ($a\rightarrow a e^{-i\omega_0t}$), we find:
\begin{eqnarray}
\frac{\partial P}{\partial t}&=&-\sum_i\frac{\partial}{\partial a_i}B_iP+\frac{1}{2}\sum_{ij}\frac{\partial^2}{\partial a_i\partial a_j}D_{ij}P,\label{FP}
\end{eqnarray}
where $\boldsymbol{a}=\left(\begin{array}{c} a\\ a^*\end{array}\right)$,   $\boldsymbol{B}=\left(\begin{array}{c}i a(\omega_0-\omega_c-2u|a|^2+i\gamma)-if \\-i a^*(\omega_0-\omega_c-2u|a|^2-i\gamma)+if\end{array}\right)$,  $D=\left(\begin{array}{cc}-2iua^2 & 0\\ 0 & 2iua^{*2}\end{array}\right)$, and $f=\sqrt{\gamma_L/\tau b}$. Eq. (\ref{FP}) is equivalent to the Langevin equation \cite{WallsMilburn}:  
\begin{eqnarray}
\partial_t\boldsymbol{a}=\boldsymbol{B}+\boldsymbol{\zeta}(t), \label{Langevin}
\end{eqnarray}
where $\zeta_i(t)$ is  delta-correlated random noise terms so that
$\langle \zeta_i(t) \zeta_j(t')\rangle =D_{ij}\delta(t-t')$.

For  small fluctuations, we can expand $a$  around the ensemble average $\langle a \rangle$, which satisfies the steady-state equation 
$|\langle a\rangle |^2[\gamma^2+(\omega_0-\omega_c-2u|\langle a\rangle|^2)^2]=|f|^2$. We introduce new real variables, $r$ and $\theta$, so that  $a=\langle a \rangle (1+r-i\theta)$ and rewrite  Eq. (\ref{Langevin}) for $\mathbf{X}=\left(\begin{array}{c} r\\ \theta \end{array}\right)$ as:
\begin{eqnarray}
\partial_t \mathbf{X}=A\mathbf{X}+\boldsymbol{\eta}(t), \label{rtheta}
\end{eqnarray}
where $A=\left(\begin{array}{cc} -\gamma & \omega_0-\omega_c-2|\langle a \rangle |^2u\\
-(\omega_0-\omega_c-6|\langle a \rangle |^2u) & -\gamma \end{array}\right)$, $\boldsymbol{\eta}=\frac{1}{2}\left(\begin{array}{c}\zeta_1/\langle a \rangle +\zeta_2/\langle a^* \rangle\\ i(\zeta_1/\langle a \rangle -\zeta_2/\langle a^* \rangle)\end{array}\right)$. For the initial condition $X(0)=0$, Eq. (\ref{rtheta}) has the solution
\begin{eqnarray}
\mathbf{X}(t)=e^{At}\int_0^te^{-At'}\boldsymbol{\eta}(t')dt', \label{X}
\end{eqnarray}
which for the spectral function $S_L^R(\omega)$ (Eq. (\ref{SSSS})) yields
\begin{eqnarray}
S_{L}^{R}(\omega)=|b|^4
\left(I_{L}^{R}\delta (\omega-\omega_0) +\frac{\gamma_L^2\gamma_{R}}{\gamma^3}F(\omega_0)\tilde{F}(\omega)\right),
\label{qS}
\end{eqnarray}
where
$F(\omega_0)$ and $\tilde{F}(\omega)$ are given below  Eq. (\ref{S}).

%\begin{thebibliography}{100}
%\bibitem{Verger} Verger, A.  Ciuti, C. \&  Carusotto, I. J.  {\it Phys. Rev. B} {\bf 73}, 193306 (2006).
%\bibitem{Jung-Tsung Shen}  Shen, Jung-Tsung \& Fan, Shanhui  {\it Phys. Rev. Lett.} {\bf 98}, 153003 (2007).
%\bibitem{WallsMilburn} Walls, D. F.  \&  Milburn, G. J. {\it Quantum Optics} (Springer-Verlag, Berlin, 1994).
%\end{thebibliography}

\bibliographystyle{apsrev}
\bibliography{strongampl}

\end{document}